# Dielectric properties and dynamical conductivity of LaTiO$_3$: From dc to optical frequencies

P. Lunkenheimer[1], T. Rudolf[1], J. Hemberger[1], A. Pimenov[1], S. Tachos[1], F. Lichtenberg[2], and A. Loidl[1]

[1] *Experimentalphysik V, Elektronische Korrelationen und Magnetismus, Institut für Physik, Universität Augsburg, 86135 Augsburg, Germany*
[2] *Experimentalphysik VI, Elektronische Korrelationen und Magnetismus, Institut für Physik, Universität Augsburg, 86135 Augsburg, Germany*

We provide a complete and detailed characterization of the temperature-dependent response to ac electrical fields of LaTiO$_3$, a Mott-Hubbard insulator close to the metal-insulator transition. We present combined dc, broadband dielectric, mm-wave, and infrared spectra of ac conductivity and dielectric constant, covering an overall frequency range of 17 decades. The dc and dielectric measurements reveal information on the semiconducting charge-transport properties of LaTiO$_3$, indicating the importance of Anderson localization, and on the dielectric response due to ionic polarization. In the infrared region, the temperature dependence of the phonon modes gives strong hints for a structural phase transition at the magnetic ordering temperature. In addition, a gap-like electronic excitation following the phonon region is analyzed in detail. We compare the results to the soft-edge behavior of the optical spectra characteristic for Mott-Hubbard insulators. Overall a consistent picture of the charge-transport mechanisms in LaTiO$_3$ emerges.

## 1. INTRODUCTION

Among the perovskite-related transition-metal oxides, that are in the focus of solid-state research since almost two decades now, LaTiO$_3$ is a prominent member and of high interest for a variety of reasons: At first, LaTiO$_3$ is a prototypical Mott-Hubbard insulator, which is just at the verge of a metal-insulator (MI) transition, becoming metallic already at very low doping levels.[1,2,3] This notion[4] was corroborated by a variety of experimental investigations,[2,3,5,6,7,8,9,10] with optical spectroscopy of electronic excitations playing a leading role.[5,6,7,8,9,10] Very recently, renewed interest in LaTiO$_3$ arose due to experimental observations[11] and theoretical interpretations[12,13] of the ground-state properties in terms of an "orbital-liquid" state, involving the dynamical quenching of the orbital moments in LaTiO$_3$. The orbital degrees of freedom are nowadays understood to play a crucial role for the electronic and magnetic properties of many transition metal oxides. While the existence of an orbital-liquid state in LaTiO$_3$ is still a matter of controversy,[11,12,14,15,16,17] it is clear that the disclosure of the orbital ground state in LaTiO$_3$ is a highly relevant and complex problem, both, experimentally and theoretically. Closely connected to this topic is also the question of a structural phase transition, which, via the JT-effect, accompanies possible orbital order and may occur simultaneously with the antiferromagnetic ordering at $T_N \approx 146$ K. Recent experimental findings indeed revealed some hints for structural anomalies at $T_N$.[16,17] Finally, LaTiO$_3$, can be expected to exhibit also interesting dielectric properties: Within the well-known La$_{1-x}$A$_x$TiO$_3$ (A = Ba or Sr) phase diagrams,[3,18] the compositions with $x = 1$ show ferroelectric (Ba) or incipient ferroelectric (Sr) behavior. While the electrical ordering in these materials usually is explained in terms of ionic displacements, recently also the importance of electronic correlations for ferroelectricity in these systems was pointed out.[19] In addition, in the framework of the Falicov-Kimball model even the occurrence of electronic ferroelectricity in correlated electron systems was proposed.[20] As LaTiO$_3$ is very close to a MI transition, irrespective of possible ferroelectric correlations, high values of the dielectric constant can be expected; however, up to now no comprehensive study of the dielectric properties of LaTiO$_3$ was performed.

Among the previous investigations of the response of LaTiO$_3$ to electromagnetic fields, so far the main emphasize has focused on the study of its electronic excitations via infrared (IR) and optical spectroscopy.[5,6,7,8,9,10] Some information was also gained on its phonon modes via far infrared (FIR) spectroscopy.[10,21,22] Most of the earlier investigations of LaTiO$_3$ in the FIR region were performed at room temperature only and often with rather low precision, revealing, e.g., only a small fraction of the 25 IR-active phonon modes, expected for orthorhombic LaTiO$_3$. The same can be said for the investigation of the region beyond the phonon modes. Also the correct stoichiometry of the samples, used in early investigations of LaTiO$_3$, sometimes seems doubtful. Investigations at lower frequencies, including the range of classical dielectric spectroscopy from Hz to MHz frequencies, are almost completely missing up to now. In ref. 23, very high values of the dielectric constant $\varepsilon' > 1000$ were reported for high temperatures, $T > 600$ K, but the intrinsic nature of these results is not clear as contact effects may also lead to apparently high dielectric constants.[24]

In the present work we provide spectra of the ac conductivity $\sigma'$ and dielectric constant $\varepsilon'$ of single-crystalline LaTiO$_3$ over an overall frequency range of 17 decades, from 10 mHz to 1000 THz. This is achieved by combining results from classical dielectric and coaxial-reflection spectroscopy, as well as from quasi-optic mm-wave and IR spectroscopy. The measurements were performed from room temperature down to 6K and for two different crystallographic directions. The samples used in this work are very close to ideal stoichiometry, especially concerning the oxygen content, where already slight deviations can lead to significant free-carrier contributions.[1]

One of the aims of the low-frequency ($\nu$ < 2 GHz) measurements, performed in this work, is to determine the intrinsic dielectric constant at frequencies below the phonon modes, which is caused by the ionic and electronic polarizability. Another important point is the clarification of the role of Anderson localization in LaTiO$_3$. While the Mott-Hubbard- or Charge-Transfer-insulator scenario nowadays is used as standard interpretation of the non-metallic behavior of various transition-metal oxides, the importance of Anderson-localization is often neglected. Via a characteristic frequency dependence of the conductivity below GHz, it was demonstrated that in many semiconducting or insulating transition-metal oxides, including high-$T_c$ related materials and colossal-magnetoresistance manganates, hopping conduction of Anderson-localized charge carriers plays an important role (see, e.g., refs. 25, 26, 27, 28, 29). It seems that even for undoped materials, small inevitable deviations from stoichiometry can generate sufficient disorder to lead to Anderson localization.[25][26][27]

Our measurements in the FIR region aim at the precise determination of the temperature-dependent phononic excitations in order to check for possible structural anomalies at $T_N$. First results have already been published in ref. 16. Furthermore, the comparison of the dielectric constant in the FIR region and the GHz region should allow to conclude for the origin – either ionic or electronic – of the intrinsic dielectric constant observed in dielectric spectroscopy. In addition, it is of interest to clarify the contribution from charge transport to the frequency-dependent conductivity in the low FIR region by comparison to the results from dielectric spectroscopy and measurements in the mm-wave range. For the description of this region, in earlier literature a background contribution from free-carrier Drude-behavior was evoked,[5][8][21][22] however, the samples used in these investigations can be suspected to have been slightly off-stoichiometric. Finally, we intend to obtain information about the gap-like excitations in the region beyond the phonon modes, including its so far only rarely investigated temperature dependence. Special emphasis will be paid to the onset of the interband conductivity. A soft-edge behavior is expected for a Mott-Hubbard insulator, while a Slater antiferromagnet would produce a sharply rising conductivity at the band edge.[30]

## 2. EXPERIMENTAL DETAILS

Single crystals of LaTiO$_3$ were prepared by floating-zone melting as described in detail in ref. 1. Some of the crystals were untwinned and these were used for the measurements. The oxygen content was determined by thermogravimetry, yielding a stoichiometry of La:Ti:O like 1:1:2.99. The x-ray pattern at room temperature revealed an orthorhombic structure (P$bnm$, $z$ = 4) with the lattice parameters $a$ = 5.633, $b$ = 5.617, $c$ = 7.915 Å. The Neel temperature of LaTiO$_3$ depends sensitively on the oxygen content. The $T_N$ = 146 K of our samples is among the highest reported in literature,[16] a fingerprint for crystals very close to ideal stoichiometry. The same can be said for the low dc-conductivity,[16] as deviations from stoichiometry drive the Mott-Hubbard insulator towards a more metallic state. For the measurements of the present work, several platelets were cut from the same melt-grown rod. From x-ray and thermal expansion measurements performed on platelets cut in the same direction,[16] the crystallographic $a$-axis was deduced to be perpendicular to the sample surface. The IR measurements were performed in reflection geometry on a polished platelet with an area of about 3x3mm and 1mm thickness, the electric field being oriented perpendicular to the $a$-axis. For the measurements from dc to 2 GHz smaller platelets were covered with two silver-paint contacts in sandwich and coplanar geometry, thus enabling measurements parallel and perpendicular to the $a$-axis.

The dc conductivity was measured with a standard four-point technique, employing an electrometer. The ac electrical response from 20 Hz to 1 MHz, was determined using an autobalance bridge (HP4284A LCR-meter). At temperatures below 60 K, additional measurements down to 10$^{-2}$ Hz were performed using a frequency-response analyzer (Novocontrol α-analyzer). In all these measurements, for cooling, the samples were mounted in a $^4$He-bath cryostat. In the radio to microwave regime, a reflectometric technique employing an impedance analyzer (Agilent 4291A, 1 MHz – 1.8 GHz) was used, the sample shorting the inner and outer conductor at the end of a coaxial line.[31] For cooling, the end of the coaxial line was connected to the cold head of a closed-cycle refrigerator, allowing measurements down to 30 K. For measurements below about 100 K, an open/short compensation has been performed at low temperatures to eliminate an influence of slight temperature-dependent changes of line and sample holder. In the IR regime the temperature-dependent reflectivity $R$ was determined using two Fourier-transform spectrometers (Bruker IFS 113v and IFS66v/S). A spectral range from 80 cm$^{-1}$ ($\approx$ 2.4 THz) to 25000 cm$^{-1}$ ($\approx$ 740 THz) was covered utilizing a suitable set of sources, beamsplitters, windows, and detectors; reference measurements were performed with a gold mirror at $\nu$ < 16000 cm$^{-1}$ and an aluminum mirror at higher frequencies. In addition, in regions of small reflectivity, the background reflectivity due to cryostat windows and unwanted reflections within the spectrometer was taken into account by making a correction using the measurement results of a zero-reflectivity sample. Cooling down to 6 K was achieved by a $^4$He-bath cryostat. Spectra of $\sigma'$ and $\varepsilon'$ were calculated from the measured reflectivity and the phase shift, which was calculated via a Kramers-Kronig (KK) transformation. The required low-frequency extrapolation was based on the measured dielectric data, while at high frequencies a $\nu^{-1.55}$ power law extrapolation was found to match the experimental data best. To check for the behavior in the frequency gap between the highest frequency of the coaxial measurements (1.8 GHz) and the lowest FIR frequency (2.4 THz), additional transmission measurements were carried out in the mm-wave region using a quasi-optic spectrometer in Mach-Zehnder configuration.[32] It allows for the measurement of transmittance and phase shift, thus avoiding the necessity of a KK transformation. Using this setup, measurements at 145 GHz were performed between 5 and 300 K.

## 3. RESULTS AND DISCUSSION

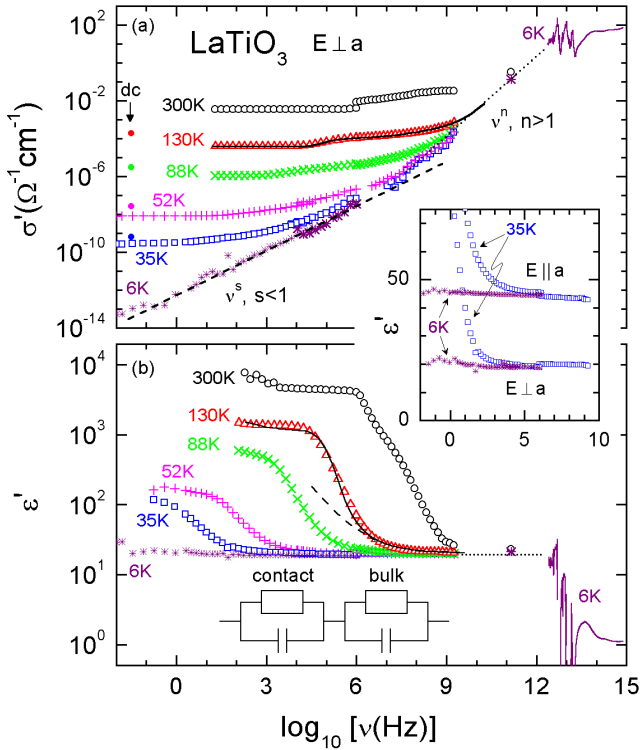

**FIG. 1.** Frequency-dependent conductivity (a) and dielectric constant (b) of LaTiO$_3$ for selected temperatures and field direction $E \perp a$ in the complete frequency range investigated. For clarity reasons, in the IR region the results are shown for 6 K only. The dashed and dotted lines in (a) indicate the UDR and SLPL, respectively, for the lowest temperatures. The dotted line in (b) indicates the intrinsic $\varepsilon_\infty$. The solid lines through the 130 K curves are the results of a fit performed simultaneously for $\sigma'(\nu)$ and $\varepsilon'(\nu)$, using the equivalent circuit indicated in (b) with the bulk contribution including the sum of UDR and SLPL behavior. The dashed line in (b) shows the intrinsic bulk contribution for 130 K. The inset compares the dielectric constant measured for $E \perp a$ and $E || a$ at two temperatures.

In Fig. 1 the spectra of the conductivity and dielectric constant of LaTiO$_3$ are shown in the complete frequency range investigated for selected temperatures and with the electric field $E \perp a$. For frequencies $\nu < 2$ GHz, measurements were performed also for $E || a$ revealing a qualitatively similar behavior.[33] The spectra show a rich variety of features that will be treated in the following chapters.

### 3.1 Electrode Effects vs. Intrinsic Contributions

In Fig. 1a, at low frequencies, a plateau shows up in $\sigma'(\nu)$, which is most pronounced for the higher temperatures. At $T \geq 35$ K, this plateau is interrupted by a small step-like increase (e.g., at about $10^5$ Hz for 130 K), which shifts towards higher frequencies with increasing temperature. It is quite smeared out at $T \leq 52$ K, due to the further increase of $\sigma'(\nu)$, as discussed below. In addition, at 300 K a significant offset between the LCR-meter and impedance-analyzer results shows up at 1 MHz. The solid circles in Fig. 1a indicate the dc conductivity, obtained from a four-point measurement. Obviously, the conductivity values detected at the lowest frequencies of the ac measurements are lower than those of the four-point measurement. This indicates significant contact contributions leading to an apparently lowered conductivity in the two-point ac-measurements. Metal-semiconductor contacts often exhibit high ohmic resistances, accompanied by a high capacitance due to the formation of a depletion layer at the metal-to-semiconductor interface.[34] They usually can be modeled by a parallel RC-circuit, connected in series to the bulk sample, thus leading to the equivalent circuit indicated in Fig. 1b.[24,27] At high frequencies, the contact resistor becomes shortened by the contact capacitance and the intrinsic bulk response is detected. This causes the smooth step-like increase of $\sigma'(\nu)$, observed at the higher temperatures. Indeed, the values reached by $\sigma'$ beyond the step agree well with the results from the dc measurement (Fig. 1a). The observed shift of the steps with temperature can be understood taking into account the temperature-dependence of the circuit's time constant, which is mainly determined by the semiconducting characteristics of the sample resistance.[24] At 6 K, the step has shifted out of the frequency window and intrinsic behavior prevails for the complete curve. The mentioned offset at 1 MHz in the 300 K spectrum mirrors the fact that two crystals with different geometry were used for the low and high-frequency measurements. Thus the differing contact resistance leads to the observed offset, showing up in the non-intrinsic region of $\sigma'(\nu)$. For the lower temperatures, the merging-frequency of both methods is located beyond the frequency of the contact-step (which has shifted to lower frequencies), i.e. in the intrinsic region and thus no offset is observed. In $\varepsilon'(\nu)$ (Fig. 1b), the high contact capacitance leads to apparently huge values of the dielectric constant in the contact-dominated region. Only at high frequencies the intrinsic $\varepsilon'$ is observed. To check for the validity of the interpretation outlined above, we have carried out a least-square fit of the spectra at 130 K using the equivalent circuit shown in Fig. 1b. To take account of the high-frequency increase of $\sigma'(\nu)$, for the bulk response an additional frequency-dependent term was included, as described in section 3.3. The results of the fit, performed simultaneously for $\sigma'$ and $\varepsilon'$, are shown as solid lines in Figs. 1a and b. A reasonable agreement of experimental data and fit can be stated. With $R_{dc} \approx 120$ k$\Omega$ and $R_c \approx 180$ k$\Omega$ the fit reveals quite similar magnitudes of dc resistance and contact resistance. In contrast, the thin depletion layer corresponds to a contact capacitance $C_c \approx 24$ pF, much larger than the intrinsic capacitance of $C_i \approx 0.16$ pF. In comparison to the fit, the step in $\varepsilon'(\nu)$ appears somewhat smeared out. This finding may be attributed to the roughness of the sample surface, leading to a distribution of time constants of the circuit. At 300 K, even an indication of two successive contact steps shows up, the origin of which is unclear.

### 3.2 Dc Conductivity

The plateau value, reached by $\sigma'(\nu)$ at frequencies beyond the contact step (e.g. at $10^6$ Hz for 130 K), can be identified by the intrinsic dc conductivity $\sigma_{dc}$. In Fig. 2, $\sigma_{dc}(T)$ determined from the $\sigma'(\nu)$ spectra is shown for both field directions. In addition, the solid line denotes the result of the dc four-point measurement with $E \perp a$, revealing a reasonable agreement with $\sigma_{dc}$ from the ac measurement. Both, an Arrhenius plot and a representation that should linearize the data for a behavior $\sigma' \sim \exp(-T_0/T)^{1/4}$ are given. The latter is predicted for variable range hopping (VRH), involving the phonon-assisted tunneling of Anderson-localized electrons or holes.[35] Obviously, the dc conductivity of LaTiO$_3$ does not exhibit simple thermally activated behavior over any significant temperature range. A temperature-dependent scattering, e.g. at spin-fluctuations when approaching the antiferromagnetic transition, can be suspected to cause this behavior. At $T_N = 146$ K (indicated by arrows), $\sigma_{dc}(T)$ exhibits a point of inflection. As already speculated in ref. 16, this anomaly may be attributed to orbital fluctuations in the vicinity of an orbital-order transition accompanying the magnetic order at $T_N$. At $T < 100$ K, the straight dashed lines demonstrate that VRH of Anderson-localized charge carriers dominates the charge transport at low temperatures. We obtain values of $T_0 = 1.0 \times 10^8$ K ($E \| a$) and $T_0 = 8.6 \times 10^7$ K ($E \perp a$), which is of similar magnitude as observed in various other transition metal oxides.[27,36,37] Interestingly, $\sigma_{dc}$ for the two field directions shows a considerable anisotropy, which changes sign at about 60 K ($1000/T \approx 17$). In ref. 16, from x-ray diffraction and thermal-expansion measurements it was deduced that in the orthorhombic structure of LaTiO$_3$ the lattice parameter $a$ is larger than $b$ and $c/\sqrt{2}$ at all temperatures $T < 300$ K. It was assumed that the detected enlargement of this difference at $T_N$ is connected with ferrodistortive orbital order of dumbbell-shaped orbitals ("p-like") within the $ab$ plane. At high temperatures, the higher $\sigma_{dc}$ for $E \| a$ documented in Fig. 2 may mirror an anisotropy in the intersite transfer matrix element that is influenced by orbital-order correlations with a dominant component in $a$-direction. In contrast at low temperatures, where tunneling prevails, the hopping probability mainly depends on the tunneling distance. Thus the smaller tunneling distance for $E \perp a$ probably leads to the detected enhancement of the conductivity in this direction.

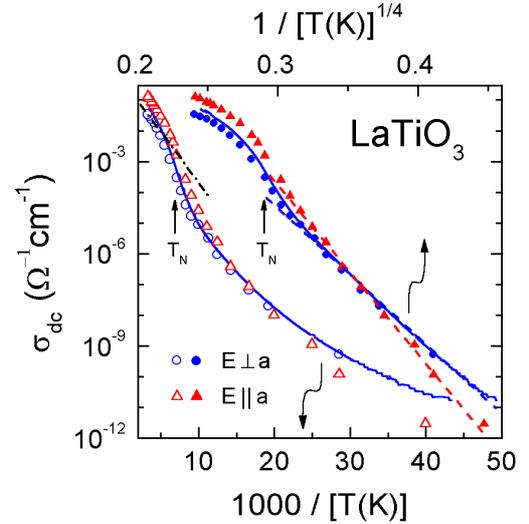

**FIG. 2.** Temperature-dependent dc conductivity of LaTiO$_3$ for both investigated field directions. The data are given in an Arrhenius (lower scale) and a VRH representation (upper scale). The solid lines are the results from a dc fourpoint measurement and the symbols were determined from the ac measurements of Fig. 1. The dashed lines indicate VRH behavior for $T < 100$ K. The dash-dotted line was calculated with an energy barrier of 65 meV, as determined from the evaluation of the gap feature in the IR region.

### 3.3 Ac Conductivity and Universal Power Laws

Following the intrinsic dc-plateau, $\sigma'(\nu)$ starts to rise continuously with increasing frequency (Fig. 1a). At the lowest temperatures, where contact step and dc plateau no longer contribute, this increase is revealed to be composed of two power laws, namely a sublinear increase $\sigma' \sim \nu^s$, $s < 1$ (dashed line), extending over about 8 frequency decades and a superlinear power law, $\sigma' \sim \nu^n$, $n > 1$ (dotted line). Taking into account the mm-wave and FIR results, the latter seems to extend well up to the low-frequency edge of the phonon region. For low temperatures, values of $s \approx 0.79$ and $n \approx 1.4$ are deduced.

A sublinear power law, $\sigma' \sim \nu^s$ corresponds to the so-called "Universal Dielectric Response" (UDR), which was demonstrated by Jonscher[34,38] to be a universal feature in the dielectric response of disordered matter. For conducting materials, it is commonly ascribed to the hopping of Anderson-localized charge carriers. There is a variety of theoretical approaches that deduce this behavior from the microscopic transport mechanisms, involving hopping over or tunneling through an energy barrier, separating different localized sites.[39] A value of $s \approx 0.8$, as found here for LaTiO$_3$ at 6 K, is typical for the ac-response of the VRH theory,[35,39] thus the ac results being in good accord with the VRH behavior revealed by the dc conductivity (Fig. 2). A prerequisite for Anderson-localization is the presence of disorder, as prevailing, e.g., in amorphous or doped semiconductors. Obviously, in pure, single-crystalline LaTiO$_3$, marginal disorder, e.g. a slight off-stoichiometry of the oxygen content or impurities at a ppm level, suffices to localize the charge carriers, so that the UDR and VRH can be observed. Also for other undoped transition-metal oxides, e.g. La$_2$CuO$_4$ (ref. 26) or LaMnO$_3$,[27] the presence of Anderson localization was revealed by ac-conductivity measurements. Via the KK transformation and the relation $\varepsilon' \sim \sigma''/\nu$, the UDR can be deduced to contribute also to the frequency-dependent dielectric constant, namely with a term $\varepsilon' \sim \nu^{s-1}$ (ref. 34). The dashed line in Fig. 1b shows the intrinsic part of the fit of the 130 K curve. It includes the UDR contribution and the dielectric constant from ionic and electronic polarization processes (sect. 3.4). Obviously, the onset of the step-like increase of $\varepsilon'(\nu)$ towards low frequencies is dominated by the intrinsic contribution from charge carrier hopping.

The origin of the superlinear power law (SLPL), observed at frequencies beyond the UDR regime and bridging the gap to the IR region is unknown up to now. In ref. 40, it is proposed that this contribution may be a universal feature of all disordered matter and indeed it is observed for such different classes of materials as doped semiconductors,[27,36,41] supercooled liquids,[42] and ionic conductors.[43] It should be noted that the observed SLPL would be difficult to explain in terms of photon-assisted hopping, which was recently invoked for the explanation of the approximately linear or even quadratic frequency dependence of $\sigma'(\nu)$, observed in various semiconductors at low temperatures and high frequencies.[44] A transition from phonon- to photon-assisted hopping should occur for $h\nu > k_B T$, a condition that clearly is not fulfilled for most of the regions where the SLPL is observed in the present work.

As seen in Fig. 1a, at the higher temperatures it is difficult to separate the UDR and the SLPL contribution to $\sigma'(\nu)$, a fact that prevents a precise determination of the UDR exponent $s$. However, even at 130 K, it is not possible to describe the very smooth transition from the dc plateau to the SLPL without invoking an UDR contribution. This was checked by omitting the UDR in the fit performed for this temperature, which led to an unsatisfactory description of the transition region. At higher temperatures, missing data in the transition region do not allow for a conclusion on the presence of a UDR contribution. As mentioned, the SLPL extends well to the phonon modes. Thus, there is no indication of a Drude-contribution of free charge carriers, in contrast to the assumptions in some earlier works.[5,8,21,22] The found SLPL is of high relevance for the low-frequency extrapolation of IR-reflection data, that is necessary for the calculation of the conductivity via the KK transformation. Obviously, the standard extrapolations often employed, namely a constant extrapolation or an extrapolation assuming the Hagen-Rubens limit, are not valid, at least in the case of LaTiO$_3$. In the present work, instead the SLPL was used.

### 3.4 Dielectric Constant Due to Ionic and Electronic Polarization

In Fig. 1b, at the high-frequency plateau of $\varepsilon'(\nu)$ an intrinsic $\varepsilon'$ of $\varepsilon_\infty = 20$ ($\pm 2$) can be read off for $E \perp a$.[45] This value matches reasonably with the results from the mm-wave experiment at 145 GHz and with the value read off at the low-frequency edge of the FIR spectrum. Interestingly, for $E \| a$ a significantly higher value of $\varepsilon_\infty = 44$ ($\pm 2$) is observed (inset of Fig. 1). In this region, contact effects, orientational polarization and charge transport no longer contribute to $\varepsilon'$ and its value is determined by the ionic and electronic polarizability only. The measured IR spectrum of $\varepsilon'$ for $E \perp a$ allows for a separation of both contributions: In Fig. 1b, $\varepsilon'(\nu)$ is seen to drop from 20 to a value of about 2 already in the phonon region thus revealing the almost completely ionic origin of the observed $\varepsilon_\infty$. Concerning the observed anisotropy, it most likely arises due to a higher ionic polarizability of the titanium ion along $a$. In ref. 17, an elongation of the basal plane of the TiO$_6$ octahedra along $a$ was deduced from diffraction measurements and assumed to imply ferrodistortive orbital order. Thus a higher polarizability in the anisotropically deformed octahedron along the $a$-direction would be expected. On the other hand, a strong electronic contribution to the higher $\varepsilon_\infty$ for $E \| a$ cannot be fully excluded. The electronic contribution is expected to become large on the approach of the insulator-metal transition from the insulating side.[46] The observed anisotropy then could indicate that LaTiO$_3$ is closer to the insulator-metal transition for $E \| a$, in accord with the higher dc conductivity for this field direction (Fig. 2). However, overall it seems unlikely that, while $\varepsilon_\infty$ for $E \perp a$ is almost completely determined by ionic polarization, the increase of $\varepsilon_\infty$ for $E \| a$ should arise from electronic polarization, which would imply that electronic polarization plays a significant role for $E \| a$ only.

At this point it seems appropriate to make some fundamental remarks on the coexistence of polar and magnetic order.[47] Compounds with an empty $d$-shell, e.g. BaTiO$_3$ (Ti$^{4+}$: $d^0$) reveal polar order and of course exhibit no long-range magnetic order. Compounds with a partly filled $d$-shell, e.g. LaTiO$_3$ (Ti$^{3+}$: $d^1$) reveal magnetic but no polar order. Obviously the partly filled $d$-orbitals hamper the strong static or dynamic ionic polarization effects that often drive polar order. However, the remaining strong anisotropy in the polarization effects, as observed in LaTiO$_3$ must strictly be correlated with the occupancy of the orbitals and the orbital order.

### 3.5 Phonon Modes

In cubic perovskites, ABO$_3$, there are three IR-active phonon modes: With increasing frequency the external, bending, and stretching modes successively appear in the spectra They are ascribed to the vibration of the BO$_6$ octahedra against the A ions, the bending of the Ti-O bond-angle, and the variation of the B-O bond length, respectively. For an orthorhombically distorted perovskite, like LaTiO$_3$, factor group analysis predicts a splitting of each of these modes into five IR-active modes and the appearance of four additional bending modes, which are inactive in the cubic case.[48,49] In addition, six new modes should arise in the IR spectra due to the orthorhombic distortion, making a total of 25 IR-active phonons. As LaTiO$_3$ is quite close to cubic, the splitting of the cubic modes and the amplitude of the new modes can be expected to be weak and indeed in earlier works, only few of the expected 25 modes could be resolved.[6,10,21,22] The only work providing temperature-dependent data in the FIR region to our knowledge is ref. 6, but unfortunately the authors do not provide any analysis of the observed phonon modes.

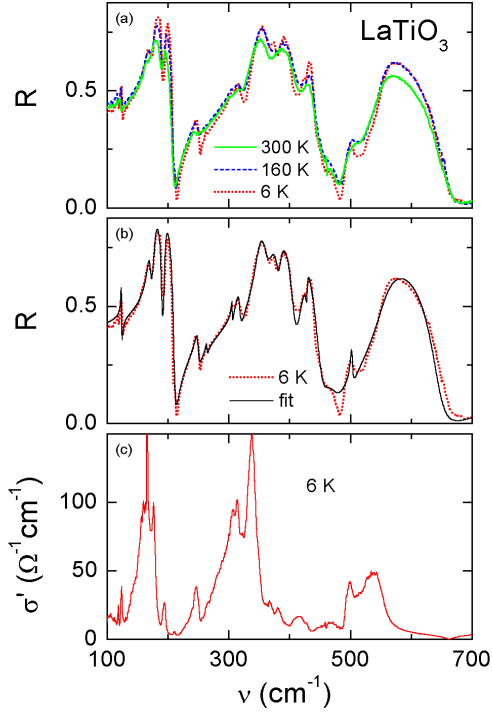

**FIG. 3.** Frequency-dependent reflectivity (a,b) and conductivity (c) of LaTiO$_3$ in the region of phonon excitations. Frame (a) shows the spectra for three selected temperatures, in (b) a fit is shown (solid line) using the sum of 17 Lorentz oscillators, and in (c) the conductivity at 6 K, deduced from a KK transformation is presented.

Figure 3 shows the FIR results of the present work with Fig 3a presenting the reflectance spectra for three selected temperatures. A large number of different modes with a significant temperature dependence of their amplitudes and frequency positions can be discerned. In Fig. 3c the conductivity at 6 K, obtained from $R$ via a KK transformation is shown. Comparing Figs. 3a and c it becomes obvious that the mode position and relative amplitude, estimated from the $R$ representation, can be misleading. For example the modes around 400 cm$^{-1}$ appear strong in the $R$ representation but reveal a low amplitude in the $\sigma$ representation. However, one has to be aware that the KK transformation can introduce some uncertainty into the conductivity spectra and often it may be preferable to analyze the raw reflectance data. The observed phonon modes can be roughly divided into three groups centered at about 170, 340, and 540 cm$^{-1}$. In accord with earlier work,[10,22] we ascribe them to the external, bending, and stretching modes of the cubic case, split due to the orthorhombic lattice distortion. In agreement with ref. 22, the center group at 340 cm$^{-1}$ seems to be split into a larger number of submodes than those at 170 and 540 cm$^{-1}$ as it is indeed expected for a bending mode.[48] In Fig. 3b a fit of the reflectance spectrum at 6 K is shown, using the sum of several Lorentz-oscillators. At 6 K, at least 17 oscillators are necessary to achieve a reasonable fit of the experimental data. At high temperatures the number of Lorentz functions was reduced, reaching 13 at room temperature, as here the modes become broadened and some of the smaller peaks are no longer resolved. In ref. 21, nine separate modes could be discerned, while in other works only the three main modes of the cubic case were analyzed.[10,22] Table 1 denotes the resonance frequencies $\nu_0$ and amplitudes $\Delta\varepsilon$ as resulting from the fits for 6 K and room temperature. While the fit in Fig. 3b looks reasonable, some details of the experimental spectrum are not quite reproduced.

| 6 K | | 300 K | |
| --- | --- | --- | --- |
| $\nu_0$(cm$^{-1}$) | $\Delta\varepsilon$ | $\nu_0$(cm$^{-1}$) | $\Delta\varepsilon$ |
| 123.5 | 0.33 | 120.5 | 0.28 |
| 169.0 | 2.41 | 166.5 | 2.27 |
| 178.6 | 3.12 | 175.8 | 3.41 |
| 195.0 | 0.68 | 191.5 | 0.37 |
| 248.4 | 0.21 | 244.5 | 0.11 |
| 264.1 | 0.03 | 255.5 | 0.02 |
| 305.6 | 0.06 | 301.5 | 0.09 |
| 316.1 | 0.46 | 314.9 | 0.55 |
| 348.4 | 3.07 | 343.9 | 3.26 |
| 368.2 | 0.76 | 375.5 | 1.31 |
| 384.3 | 0.14 | 421.5 | 0.15 |
| 388.0 | 0.08 | 506.5 | 0.01 |
| 419.0 | 0.35 | 549.5 | 1.36 |
| 429.2 | 0.01 | | |
| 467.7 | 0.09 | | |
| 502.1 | 0.04 | | |
| 552.9 | 1.16 | | |

TABLE 1. Resonance frequencies $\nu_0$ and amplitudes $\Delta\varepsilon$ from a fit of the reflectance in the phonon region using a sum of Lorentz oscillators for 6 and 300 K.

This drawback may be overcome by using more oscillators in the fits, but already with 17 oscillators the problem arises that the parameters are correlated to some extent. The main aim of the FIR investigations of the present work is the search for possible small anomalies in the temperature dependence of the phonon modes, giving hints to a possible structural phase transition. The mentioned correlation of parameters, together with the uncertainties due to the KK transformation, renders it difficult to gain such information from the fits. Namely these problems reduce the precision of the parameters to an extent that the small changes at a possible phase transition may not be caught. Thus it seems most appropriate to perform a direct inspection of the raw reflection and conductivity data for such anomalies. One example is given in Fig. 4, where a detailed view of the mode at 340 cm$^{-1}$ is given for various temperatures (for clarity reasons, spectra are shown only for selected temperatures). The inset shows the temperature-dependent frequency of the peak maximum, $\nu_0$, read off from the $\sigma'(\nu)$ curves. While at low temperatures, up to about 150 K, $\nu_0$ is effectively constant, at higher temperatures a significant shift towards lower frequencies is observed. A similar characteristic is revealed by the peaks close to 175, 195, and 535 cm$^{-1}$ (not shown) with the transition from constant to temperature-dependent $\nu_0(T)$ always lying in the range of 110-150 K. One may speculate that this behavior mirrors the fact, that the lattice constants are nearly constant at $T < 100$ K, but $b$ increases considerable at $T > 150$ K, both $a$ and $b$ exhibiting a smeared-out anomaly close to $T_N$.[16] This finding was interpreted as evidence for a structural phase transition at $T_N$.[16] Then the variation of the orthorhombic lattice distortion at high temperatures, inducing a variation of splitting of the cubic modes, may lead to the observed shift of the phonon frequencies at $T > T_N$. This implies that the modes at 175, 195, 340, and 535 cm$^{-1}$, which show evidence of this anomaly, belong to the split cubic modes, which seems reasonable (compare Fig. 3c). However, one should be aware that anharmonic effects would lead to quite a similar temperature dependence of phonon modes, namely an increase of $\nu_0$ towards lower temperatures and a saturation at low temperatures.

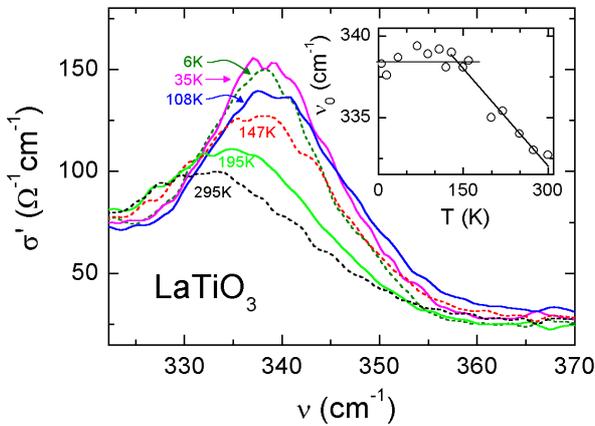

**FIG. 4.** Frequency-dependent conductivity of LaTiO$_3$ around the phonon mode at 340 cm$^{-1}$ for selected temperatures. The inset shows the temperature-dependence of the peak position $\nu_0$, with the lines indicating a change of characteristics close to $T_N$.

Fig. 5 demonstrates further anomalies of the phonon modes close to $T_N$ that occur around 350-430 cm$^{-1}$. In this region already Crandles et al.,[21] despite a reduced resolution of the spectra, found indications for two modes at 380 and 411 cm$^{-1}$, which they considered as the $F_{2u}$ modes of the cubic lattice, i.e. additional bending modes becoming IR active in the orthorhombic case. As the phonon resonances in this region have very low amplitude (cf. Fig. 3c), in order to avoid any uncertainty due to the KK transformation, in Fig. 5 the reflection raw data are plotted. The spectra reveal the appearance of two additional peaks at 373 and 424 cm$^{-1}$, showing up just at temperatures below $T_N$. This finding gives further strong evidence for a structural anomaly accompanying the magnetic phase transition at $T_N$ as promoted in refs. 16 and 17. The increase of the difference of lattice constants $a$ and $b$ at $T_N$, found in ref. 16, can be interpreted as an increase of orthorhombic distortion, which at $T < T_N$ gives rise to an increase of the amplitude of modes, invisible in the less distorted case at $T > T_N$.

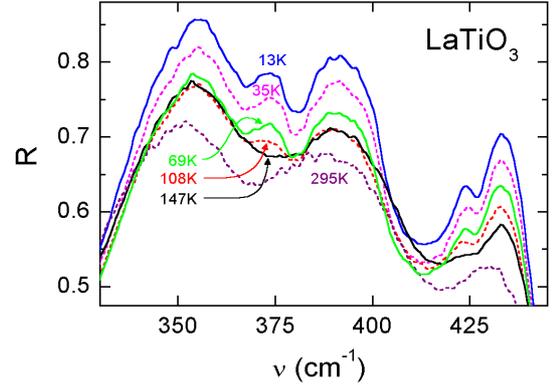

**FIG. 5.** Frequency-dependent reflectivity of LaTiO$_3$ at the group of phonon modes around 400 cm$^{-1}$ for various temperatures.

### 3.6 Electronic Excitations

#### 3.6.1 Dynamic conductivity

Figure 6 shows the conductivity at the highest frequencies investigated, including the last phonon mode. At frequencies beyond the phononic contributions, $\sigma'(\nu)$ exhibits the typical signature of an electronic excitation across an energy gap. Namely, after dropping to very low values, below the detection limit, it shows a steep increase with an onset close to 700 cm$^{-1}$, followed by a saturation above about 3000 cm$^{-1}$. In accord with previous interpretations,[6 7 8 9 10] this gaplike feature can be ascribed to a transition between the lower and upper Hubbard bands in the typical Mott-Hubbard insulator LaTiO$_3$. The inset of Fig. 6 shows the results of Fujishima et al.,[5] which extend beyond the highest frequencies investigated in the present work.[50] They reveal another much stronger excitation with an onset at about $3 \times 10^4$ cm$^{-1}$, which was interpreted as charge-transfer transition between the O 2$p$ and Ti 3$d$ bands.[5 7 10] In refs. 6, 7, 9, and 10, a quantitative evaluation of the Mott-Hubbard gap energy $E_g$ was performed by using a simple linear extrapolation in the $\sigma'(\nu)$ plot. Room-temperature values of $E_g = 20$ meV,[6] 0.1 eV,[7 10] and 0.2 eV[9] were reported, the variation reflecting uncertainties due to the choice of the section of the spectrum used for the extrapolation and possible differences of stoichiometry of the samples. Only in ref. 6, temperature-dependent measurements were reported for four different temperatures, the gap energies varying between about 20 meV at 300 K and 50 meV at 20 K. To obtain more precise information on the temperature-dependent gap energies and to use a less ambiguous way of extrapolation, here we analyzed the spectra in the gap region within simple models based on Fermi's golden rule. As noted in textbooks,[35,51] assuming parabolic and three-dimensional bands, the frequency-dependent conductivity in the gap region can be calculated. It is predicted to show significantly different characteristics for the cases of direct and indirect and for quantum-mechanically allowed and forbidden transitions. Namely, a plot of $(\sigma' \times \nu)^n$ with $n = 2, 2/3, 1/2$, and $1/3$ should linearize for direct allowed, direct forbidden, indirect allowed, and indirect forbidden transitions, respectively, and an extrapolation to an ordinate value of zero should allow to read off the gap energy. Checking these representations reveals the case of a direct forbidden transition best describing the experimental data.[33] Fig. 7 shows the corresponding plot for 6 K. The finding of the forbidden nature of the observed transition is reasonable following the common notion[6,7,8,9,10] that it occurs between the lower and upper Hubbard bands. Making the assumption that the transition takes place within the $t_{2g}$ band, split by the Hubbard energy $U$, the selection rule $\Delta l = \pm 1$ would be broken. However, this argument possibly is too naive for strongly correlated electron systems. A fit to the data analyzing the results up to 2000 cm$^{-1}$ yields $E_g = 570$ cm$^{-1} \approx 71$ meV, a rather small value compared to the Hubbard $U$ that is usually assumed to be of the order of few eV.

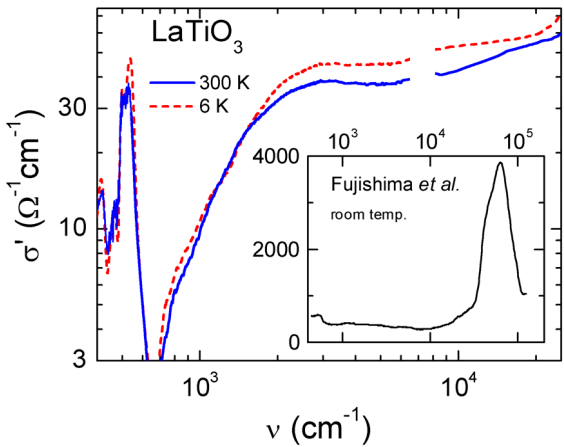

**FIG. 6.** Frequency-dependent conductivity of LaTiO$_3$ at frequencies beyond the phonon modes for two temperatures. The inset shows the results at room temperature reported by Fujishima et al.[5]

### 3.6.2 Energy gap and "soft-edge" behavior

In the upper inset of Fig. 7, a closer view of the region close to the gap energy is given for three temperatures, revealing significant temperature dependence. To investigate the temperature dependence of the gap, we analyzed the experimental results using $\sigma' \sim (\nu - E_g)^{3/2}/\nu$, which describes direct forbidden transitions.[51] It is interesting to compare this frequency dependence of the conductivity derived for optical transitions in conventional semiconductors with predictions for strongly correlated electron systems. In a classical paper on the optical energy gap in V$_2$O$_3$,[30] a "soft-edge" behavior has been reported with $\sigma' \sim (\nu - E_g)^{3/2}$ and has been described as a hallmark of strong electron correlations. Specifically it has been pointed out that this behavior clearly differs from the conductivity of a Slater antiferromagnet, which would produce a conductivity that rises sharply with $\nu^{1/2}$. One should note that for $\nu - E_g \ll E_g$, $(\nu - E_g)^{3/2}/\nu$ is approximately equal to $(\nu - E_g)^{3/2}/E_g$. Thus for frequencies just above the onset of the interband transition, the conductivity predicted for an electronically correlated system is identical to that for a direct-forbidden transition in a canonical semiconductor and deviations are expected for higher frequencies only.

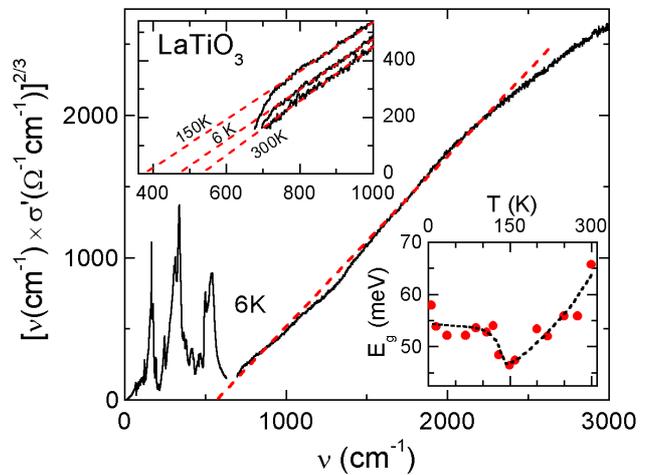

**FIG. 7.** Frequency-dependent conductivity of LaTiO$_3$ at 6 K, plotted in a way to linearize the data in the region of the electronic gap excitation for the case of a direct forbidden transition. The upper inset demonstrates the extrapolation, performed to determine the gap energies, for three selected temperatures. The lower inset shows the temperature-dependent gap energy; the line is drawn to guide the eye.

The lower inset of Fig. 7 shows $E_g(T)$ determined assuming a direct forbidden transition using the experimental results below 1000 cm$^{-1}$ (see solid lines in the upper inset). The gap energy reveals a significant temperature dependence: On decreasing temperature it decreases from about 65 meV at room temperature to 45 meV at $T_N$, where it undergoes a step-like increase to a low-temperature value of 55 meV. The increase at $T_N$ could be understood in terms of the magnetic exchange energy $J$, which is of the order of $T_N$. However, the decrease below room temperature can only be understood in terms of structural effects, e.g. lattice expansion or dynamical Jahn-Teller effects. And hence, also the increase of $T_N$ could be due to structural effects only. In this context we would like to recall that an orbital-ordering transition has be reported to occur at $T_N$.[16] In their investigation of the Mott-Hubbard gap in the system La$_x$Y$_{1-x}$TiO$_3$, Okimoto et al.[9] found a systematic increase of $E_g$ with Y-content. From calculations using a tight binding model they concluded that this behavior is due to a decrease of the one-electron bandwidth induced by an increase of orthorhombic distortion. Thus the sudden increase of $E_g$ when entering the antiferromagnetic state, observed in the lower inset of Fig. 7, may also be ascribed to the increase of orthorhombicity below $T_N$.[16]

The measured dc-conductivity (Fig. 2) allows us to clarify the importance of the observed gap for the charge transport in LaTiO$_3$. The dash-dotted line in Fig. 2 indicates thermally activated behavior, $\sigma' \sim \exp[-E_g/(k_BT)]$ with $E_g$ = 65 meV, the value determined from the optic experiments at room temperature. The experimental data close to room temperature indeed can be described in this way, but at lower temperatures, where according to the optical experiments the energy barrier should decrease, the experimental $\sigma_{dc}(T)$ shows a transition to a steeper slope in the Arrhenius plot, thus suggesting an increase of the effective energy barrier. However, one has to be aware that for the deduced relatively low energy barriers, much more carriers are excited across the gap than in usual semiconductors and the number of charge carriers may not be the sole factor determining $\sigma_{dc}(T)$. Instead, as speculated in chapter 3.2, temperature-dependent scattering processes at magnetic or orbital fluctuations, successively increasing on the approach towards $T_N$, may explain the observed reduction of $\sigma_{dc}$. Below $T_N$, the fluctuations become suppressed, leading to the reduction of slope in the Arrhenius plot and finally, below about 100 K, tunneling processes become the dominant charge transport process.

### 3.6.3 Comparison with theoretical predictions

In recent years a number of calculations of the electronic band structure of LaTiO$_3$ have appeared (e.g. refs. 52 53 54). Here we focus on the most recent calculations using the *ab initio* computational scheme LDA + DMFT (LDA: local-density approximation, DMFT: dynamical mean-field theory),[54] which currently is the most advanced tool for the theoretical investigation of materials with strong electronic correlations. These calculations, performed for doped LaTiO$_3$, concentrated on doping levels just on the metallic side of the metal-to-insulator transition (MIT). As pure LaTiO$_3$ is close to the MIT we take the characteristic energy scales from these calculations. The results of LDA (upper frame) and LDA+DMFT (lower frame) are schematically shown in Fig. 8. Here the electronic densities of state are schematically indicated, neglecting all the fine structure and only keeping the characteristic energies - the charge transfer energy $\Delta$, the Coulomb repulsion $U$, and the band widths. In addition, the quasiparticle peak appearing at the Fermi level within the LDA+DMFT approach for doped, metallic LaTiO$_3$ was omitted to account for the non-metallic state of undoped LaTiO$_3$.

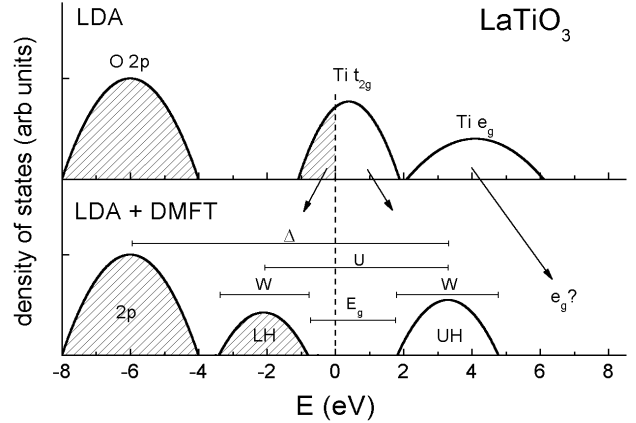

**FIG. 8.** Highly schematic representation of the density of states of LaTiO$_3$, based on the results from theoretical calculations using the LDA (upper frame) and LDA+DMFT approaches (lower frame).[54] The latter was calculated at $T$ = 1000 K for a metallic system close to the MIT taking only the $t_{2g}$ states into account and using a cubic structure with the same volume as the real orthorhombic crystal.[54] Here, for insulating LaTiO$_3$ we plotted the upper and lower Hubbard bands, leaving out the quasiparticle peak.

If we compare this calculated electronic structure with the experimental results as shown in Fig. 6, we find that the experimentally observed charge-transfer energy, $\Delta \approx 8.5$ eV is in good agreement with the calculations. However, it seems almost impossible to derive an experimental value of the Coulomb energy $U$, as in $\sigma'(\nu)$ no well-resolved peak is observed corresponding to an excitation from the lower to the upper Hubbard band. Assuming that $U$ is defined by the shoulder in $\sigma'(\nu)$ close to 3000 cm$^{-1}$, we would end up with a value of $U$ of the order 0.5 eV, much too low when compared to the *ab initio* calculations,[54] yielding values of $U$ between 4 – 5 eV. The same problem arises when comparing the experimentally observed gap $E_g$ with the theoretical values, which is of the order $E_g = U - W \approx 2$ eV, while experimentally we observe a gap well below 0.1 eV, even in highly stoichiometric LaTiO$_3$, which seems to be the most insulating crystal investigated so far. Using the existing models, $E_g < 0.1$ eV can only be explained when $U \approx W$. How can this discrepancy be explained? We note first that the calculations have been performed for a metallic sample. However, we believe that the main parameters, e.g. $\Delta$, $U$, and $W$, will not change considerably when moving to the insulating side of the MIT. One explanation could be that the symmetric densities of states of the lower and upper Hubbard band have tails towards the Fermi energy. Due to their multiplett structure they possibly could be also rather asymmetric. A further problem may arise with the empty $e_g$ bands. Their exact position is unclear and they probably mix with the upper Hubbard band. Thus electronic transitions from the lower Hubbard band to the $e_g$ band may well contribute to the dynamic conductivity below the charge transfer peak. From this it is clear that realistic LDA+DMFT calculations of the dynamic conductivity in the insulating state are highly warranted.

## 4. SUMMARY AND CONCLUSIONS

In the preceding chapters, results from extreme broadband dielectric and infrared measurements of $LaTiO_3$, performed for various temperatures and two crystallographic directions, were presented. For the first time, a thorough investigation of the dielectric and charge transport properties in the region from dc to GHz frequencies was performed. In the IR region, the phonon modes and their temperature-dependence was investigated with unprecedented precision and the gap-like feature beyond the last phonon mode was analyzed in detail. From these measurements, overall a wealth of information on charge transport, dielectric properties, and phase-transitions in $LaTiO_3$ is gained:

From the dc results and the IR measurements in the gap region, a consistent picture of the charge transport processes in $LaTiO_3$ evolves: At temperatures above about 100 K, charge transport is governed by thermally excited charge carriers that are excited over a rather small energy gap of the order of 50 meV between lower and upper Mott-Hubbard band. Due to the small gap value, the temperature dependence of $\sigma_{dc}$ is dominated by temperature-dependent scattering processes at magnetic and/or orbital fluctuations that become maximal at $T_N$. Below about 100 K, Variable Range Hopping, i.e. phonon-assisted tunneling of Anderson-localized charge carriers dominates the dc transport. Most probably they arise from defect states due to impurities or lattice defects. In full accord with these findings, also the ac conductivity shows the clear signature of hopping conduction of Anderson-localized charge carriers. Depending on temperature, in the region from 10 MHz up to IR frequencies, a superlinear power law is observed, which seems to be a universal feature of all disordered matter[40] and is important for the low-frequency extrapolation of the IR data. The dc conductivity shows a considerable anisotropy up to nearly a decade, which changes sign close to the transition from thermally activated band conduction to tunneling transport. It can be understood taking into account the anisotropy induced by the orthorhombic lattice distortion.

The intrinsic dielectric constant of $LaTiO_3$ was determined as $\varepsilon_\infty \approx 20$ for $E\perp a$ and $\varepsilon_\infty \approx 44$ for $E\|a$. These values are relatively high, when compared with conventional semiconductors, but certainly do not indicate any ferroelectric correlations. From the FIR results it is concluded that the $\varepsilon_\infty \approx 20$ for $E\perp a$ is almost completely due to ionic polarization and any electronic contributions are nearly negligible. The found anisotropy of $\varepsilon_\infty$ most probably is related to the orthorhombic distortions, enhancing the ionic polarizability along the $a$-direction.

From the detailed investigation of the temperature-dependent phonon response, strong indications for a structural phase transition, accompanying the magnetic transition at $T_N$ are obtained. This notion is further supported by an anomaly at $T_N$ found in the temperature-dependent electronic gap energy, which can be attributed to the increase of the orthorhombic lattice distortion below $T_N$. In addition, the phase transition also influences the dc transport, via an increased scattering rate, as described above. As promoted already in ref. 16, these findings speak in favor of orbital order in $LaTiO_3$ below $T_N$ and thus are at variance with the formation of an orbital liquid state.

Finally, from the optical conductivity we deduce a band gap of the order of 50 meV at low temperatures, revealing a significant temperature dependence. From a detailed analysis we conclude that the transition is direct and forbidden. Close to the gap energy this frequency dependence is identical to the soft-edge behavior, identified in $V_2O_3$ (ref. 30) and indicative for strong electron correlations.

**Acknowledgements**

We are grateful to T. Kopp, Th. Pruschke, and D. Vollhardt for stimulating discussions. This work was partly supported by the BMBF via VDI/EKM 13N6917 and 13N6918 and by the Deutsche Forschungsgemeinschaft via the Sonderforschungsbereich SFB 484 (Augsburg).